# Advancing Food Nanotoxicology with Microphysiological Systems: Rebalancing the Risk/Benefit Ratio Toward Safer Nano-Enabled Food Innovations

Georges Dubourg

Corresponding author: Georges Dubourg

University of Novi Sad, Center for Sensor Technologies, Biosense Institute, Dr Zorana Đinđića 1, 21000 Novi Sad, Serbia

**Keywords:** Food, Nanomaterial, Nanotoxicity, Microphysiological System

**Highlights**

- Highlights the risk–benefit imbalance of nanofoods linked to potential toxicity
- Positions MPS as a promising tool for food nanotoxicology assessment
- Evaluates performance criteria critical for gut modeling in global MPS advances
- Assesses the gap between general MPS advances and their application in nanotoxicology
- Proposes strategic steps to translate MPS into nanofood safety assessment

**Graphical Abstract**

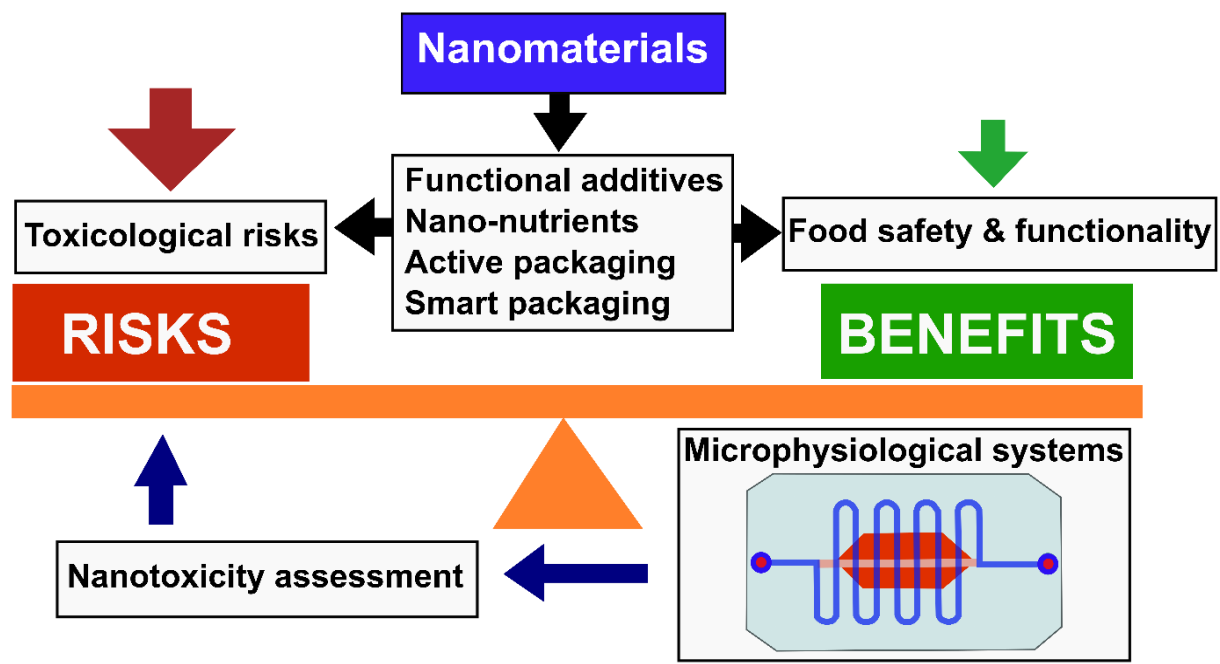

Abstract

**Background**
Incorporating nanomaterials into food products provides key benefits, including extended shelf life, improved safety, and enhanced quality and texture. These innovations could help tackle major challenges in modern food systems, such as reducing waste and enhancing food quality and safety. However, potential toxicity remains a concern, compounded by the lack of physiologically relevant models for assessing ingested nanomaterials. Traditional in vitro and in vivo approaches often fail to mimic gastrointestinal complexity, resulting in inconsistent and non-predictive nanotoxicity data that hinder accurate risk assessment of nano-enabled foods.

**Scope and approach**

To address this gap, this review evaluates the potential of microphysiological systems (MPS), particularly gut-targeted MPS, for modeling gastrointestinal nanoparticle exposure. It examines how MPS technologies replicate key physiological processes relevant to food-specific risk assessment, including intestinal barrier function, microbiota–immune interactions, and gut–organ communication. A comparative analysis of technological advances and their applications in nanotoxicology explores how MPS can be better adapted for nanofood safety evaluation.

**Key findings**

MPS technology offers a powerful tool for modeling the gastrointestinal tract by replicating key physiological features such as barrier function, microbiome-immune interactions, and gut–organ axes. However, its application in food nanotoxicology remains limited, and its full potential, particularly for microbiome–immune integration, has yet to be realized. Key challenges include sustaining long-term co-cultures, simulating digestion, and modeling nanoparticle biotransformation. Additionally, standardization, reproducibility, and regulatory alignment hinder broader adoption. This review outlines strategies to address these barriers, including AI integration and multidisciplinary frameworks to support MPS validation and regulatory uptake in nanofood safety.

**Introduction**

Rapid population growth, combined with the staggering loss or waste of nearly one-third of all food produced globally (Gonçalves et al., 2025), places unprecedented strain on the global food system, which in turn challenges its ability to ensure long-term sustainability and food safety. In this context, nanotechnology has emerged as a promising approach to address these real-world challenges. Nanomaterials, with their unique physicochemical properties such as high surface-area-to-volume ratios, tunable solubility, and enhanced reactivity, enable the development of active and intelligent packaging, as well as functional food additives (Gupta et al., 2024; Sahani and Sharma, 2020). These applications can extend the shelf life of food, boost nutrient bioavailability (Chen et al., 2023; Moradi et al., 2022), and minimize food waste, thereby supporting a more efficient and resilient food supply chain ( Herrera-Rivera et al., 2024; Dubourg et al., 2024).

Nevertheless, nanomaterials should be regarded as a double-edged sword: while they offer clear benefits in food applications, their ingestion raises significant safety and health concerns that must be carefully weighed. Their complex behavior and interactions with biological systems at the nanoscale remain poorly understood, highlighting the need for comprehensive nanotoxicity assessments to better evaluate the risks associated with their ingestion (Wang et al., 2021). To fully realize the potential of nano-enabled food products, it is essential to generate robust toxicity data alongside a clear identification of potential health risks, as these are key to ensuring a well-informed balance between innovation and consumer safety (EFSA Scientific Committee, 2021a, 2021b; Schoonjans et al., 2023).

Assessing the toxicological impact of nanomaterials remains a major challenge, as current studies often fail to produce robust data and show discrepancies across findings, leading to continued uncertainty about the toxicity of nano-enabled foods (Sohal et al., 2018; Xavier et al., 2021). A clear example is titanium dioxide ($TiO_2$), long considered safe and widely used as the food additive E171. In France, the share of food products containing $TiO_2$ dropped from 68% in 2018 to just 17% by 2021, reflecting the impact of regulatory scrutiny (Bucher et al., 2024). Following several toxicity re-evaluations, France banned its use in 2020 under the precautionary principle, followed by an EU-wide ban in 2022, whereas Japan reaffirmed its safety in 2023 (Dand et al., 2025). A key obstacle to reliable nanotoxicological data is the reliance on traditional models, such as traditional in vitro and animal studies, which often fail to replicate human physiology. This limitation is further compounded by the dynamic nature of nanomaterials, which can undergo aggregation, dissolution, and surface modifications, processes that complicate hazard identification and exposure assessment (Sohal et al., 2018; Xavier et al., 2021; Usmani et al., 2024).

In response to these limitations, there is growing momentum behind New Approach Methodologies (NAMs) that offer more human-relevant and mechanism-based toxicological assessments. These approaches support integrated evaluation of key endpoints such as genotoxicity, oxidative stress, barrier integrity, and inflammation, with improved predictive accuracy (Usmani et al., 2024; Haase et al., 2024, Cattaneo et al., 2023). Among these emerging tools, microphysiological systems (MPS), also known as organ-on-a-chip (OoC), stand out for their ability to replicate complex human tissue microenvironments under controlled, dynamic conditions (Avula et al., 2024). Although MPS have been extensively developed for pharmaceutical applications (Vulto et al., 2021), their use in nanotoxicology remains relatively underexplored. Existing reviews have generally addressed their potential from a broad toxicological perspective but have not specifically examined the applicability of MPS for assessing the toxicity of ingested food-related nanomaterials (Ashammakhi et al., 2020; Lu and Radisic, 2021).

In light of this, the present review explores how MPS can be repurposed to assess nanotoxicity from ingested nanomaterials in food products. The review begins by weighing both sides of the scale: the functional benefits of nanomaterials in food products against their potential toxicological risks. This dual perspective highlights the need for more accurate assessment tools and points to the promise of MPS-based models. The discussion then shifts to the capabilities of MPS, particularly gut-on-a-chip (GOC) platforms, to replicate key physiological features that enable quantification of intestinal absorption and metabolism of nanomaterials, including a dynamic microenvironment, a functional intestinal barrier, interactions with the gut microbiota and immune system, and interconnections along the gut–organ axis. Afterwards, selected case studies highlighting current applications of GOC in nanotoxicity research are examined and compared within a global context to identify key gaps, areas for improvement, and opportunities for broader adoption.

1. **Nanomaterials in the agrifood sector: weigh up the pros and cons**
1.1. **The Role of Nanomaterials in Modern Food Production**

Nanomaterials are used across the food production chain to enhance nutrition, sensory qualities, safety, and shelf life. They may be added directly, accumulate in food, or migrate from packaging (Xavier et al., 2021; Schoonjans et al., 2023). This section briefly reviews recent literature on their roles in improving food quality, safety, and functionality.

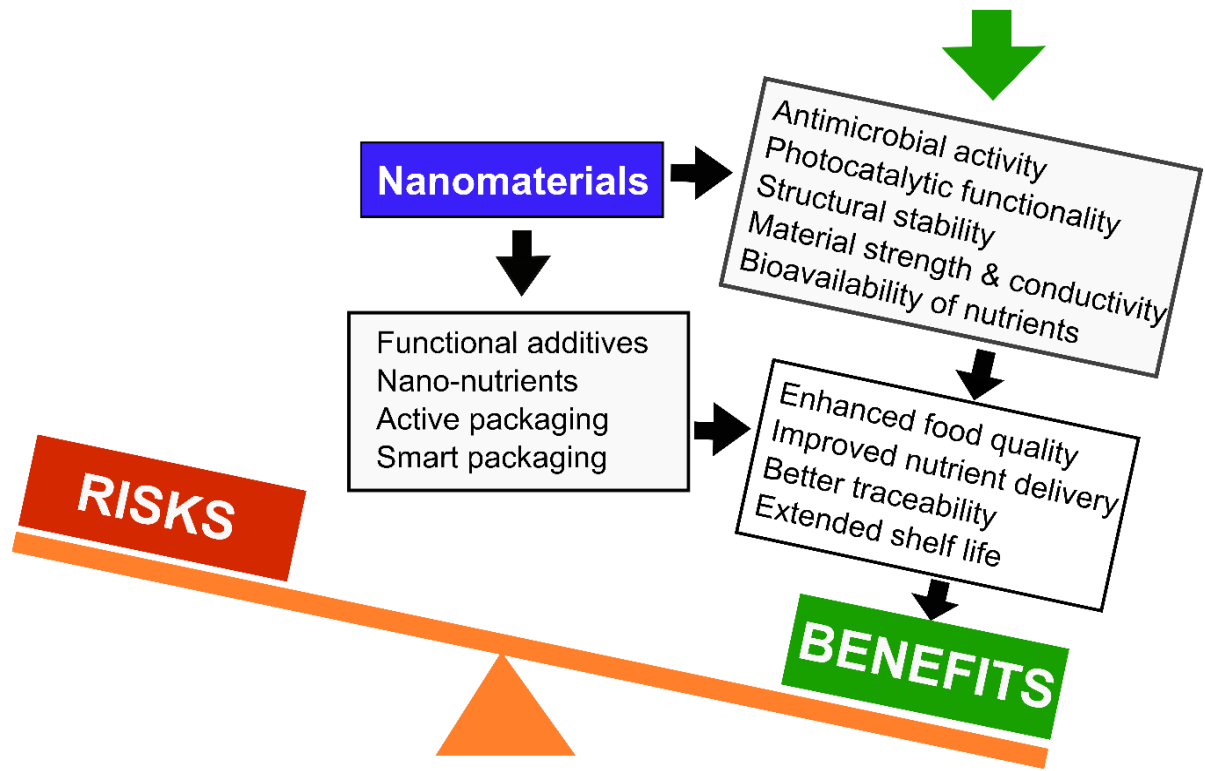

*Figure 1.* Schematic balance highlighting the benefits of nanomaterials in food-related applications.

One of the earliest applications of nanomaterials in food is their use as additives to improve quality, safety, and nutrition. Their unique properties, such as increased surface area, solubility, and reactivity, enable functions like targeted nutrient delivery, improved dispersibility, and controlled release of active compounds, surpassing conventional ingredients (Sahani and Sharma, 2020). A prominent example is titanium dioxide ($TiO_2$, E171), widely used as a whitening agent in confections, sauces, and dairy products. At the nanoscale, $TiO_2$ delivers enhanced brightness and more uniform coverage. Another widely applied nanomaterial is silicon dioxide ($SiO_2$, E551), primarily used as an anti-caking agent in powdered products such as instant coffee, soup mixes, and seasoning blends. $SiO_2$ NPs improve flowability and moisture resistance, maintaining product stability and texture (Moradi et al., 2022). Similarly, iron oxide NPs ($Fe_2O_3$, E172) serve as both colorants and nutritional supplements. Their nanoparticulate form enhances color intensity and improves iron absorption with reduced metallic aftertaste, making them ideal for fortification (Chavarría-Fernández et al., 2024; Chen et al., 2023; Mcclements & Xiao, 2017). Magnesium oxide (MgO, E530) further exemplifies the multifunctionality of nano-additives. Used in dairy and processed foods, MgO functions as an anti-caking agent, pH regulator, and carrier for flavor compounds, improving shelf life and sensory quality (Moradi et al., 2022). Calcium phosphate NPs (E341), widely included in flour and dairy formulations, contribute to better texture and processability while serving as bioavailable sources of calcium and phosphate. Notably, their behavior closely resembles naturally occurring nanoclusters in milk, supporting their safety and digestibility (Enax et al., 2022). Beyond functional improvements, some nanomaterials offer biological advantages. Silver (Ag, E174) and gold (Au, E175) NPs are traditionally used for decorative purposes in luxury confectionery, but silver NPs (AgNPs) are also explored for their antimicrobial potential. These particles show inhibitory effects against pathogens such as E. coli and S. aureus, suggesting future roles in improving food safety (Chen et al., 2023). Expanding the scope of nutritionally active nano-additives, zinc oxide (ZnO) and copper-based NPs are under investigation for their antimicrobial properties and mineral supplementation potential. A novel addition to this field is nanoselenium (SeNPs), gaining attention due to its higher bioavailability and lower toxicity compared to traditional selenium supplements, making it attractive for antioxidant and immune-supportive use in functional foods (Chen et al., 2023; Mcclements & Xiao, 2017; Moradi et al., 2022). Beyond their direct use in food, nanomaterials play a critical role in packaging by enhancing preservation, preventing contamination, and promoting sustainability. They improve the mechanical strength, flexibility, thermal stability, and barrier properties of packaging materials, as reviewed in recent studies (Nikolić et al., 2021; Prakash et al., 2024). For instance, nanoclays are recognized for significantly reducing gas permeability, thereby improving the shelf life of packaged food (Prakash et al., 2024). AgNPs provide antimicrobial properties, inhibiting bacterial growth on packaging surfaces, an effect emphasized in reviews by Chandra et al., 2024 and Nikolić et al., 2021. Additionally, ZnO and $TiO_2$ offer UV shielding, which helps prevent photodegradation of light-sensitive foods (Herrera-Rivera et al., 2024; Nikolić et al., 2021). Reinforcement with cellulose nanofibers has been proposed to improve the durability and environmental profile of biopolymer-based packaging (Prakash et al., 2024). Moreover, nanomaterials enable active packaging systems that go beyond passive containment to interact with food or its environment to prevent spoilage (Herrera-Rivera et al., 2024; Jafarzadeh et al., 2024). For instance, inorganic NPs such as Ag, ZnO, and $TiO_2$ have been incorporated into films to provide slow-release antimicrobial or antioxidant activity, enhancing food safety and extending shelf life. Silicate clays, reviewed by Jafarzadeh et al., 2024, are also noted for enhancing gas barrier properties, mitigating spoilage due to oxygen or ethylene exposure. These innovations are especially impactful for packaging perishable products like dairy, meats, and fresh produce, where microbial control and oxidation are major concerns.

The impact of nanotechnology in packaging extends beyond barrier enhancement to the development of nanosensors and nanodevices, emerging as powerful tools for real-time food quality monitoring. These technologies offer heightened sensitivity, selectivity, and responsiveness in detecting spoilage indicators, contaminants, or nutrient levels within complex food matrices. Leveraging the high surface area, tunable conductivity, and catalytic reactivity of nanomaterials, nanosensors can detect even trace levels of volatile organic compounds (VOCs), pH changes, and microbial metabolites indicative of spoilage or contamination (Sharma et al., 2023). Among sensor types, electrochemical gas sensors have gained prominence for noninvasive integration into intelligent packaging. These sensors detect resistance or potential changes upon interaction with spoilage gases like ammonia, hydrogen sulfide, carbon dioxide, and ethylene-byproducts of microbial metabolism and degradation in

protein- and produce-based foods (Nami et al., 2024, Dubourg et al., 2024). pH sensors and time-temperature indicators (TTIs) further support shelf-life monitoring by tracking acidity shifts and thermal exposure that accelerate spoilage (Siddiqui et al., 2022). The performance of these nanosensors depends on the nanomaterial used. Two-dimensional nanomaterials, such as graphene, MXenes, and layered silicate clays, are increasingly utilized due to their exceptional electrical conductivity, mechanical integrity, and surface sensitivity. Their capacity to immobilize biomolecules and facilitate rapid electron transfer makes them suitable for pH sensing and VOC detection (Sharma et al., 2023). Carbon nanotubes (CNTs) also contribute significantly to chemiresistive sensor platforms due to their high aspect ratio and conductivity. Metal and metal oxide nanomaterials are central to intelligent packaging owing to their dual role as sensing and antimicrobial agents. ZnO, for example, is widely employed in chemiresistive and colorimetric sensors for detecting spoilage gases such as ammonia and $H_2S$ while inhibiting bacterial growth. $TiO_2$, with photocatalytic properties, supports oxidation detection and contaminant degradation under UV exposure. Iron oxides ($Fe_2O_3$, $Fe_3O_4$) serve in magnetic biosensors and can be functionalized for specific contaminant detection, while silver oxide ($Ag_2O$) contributes bacteriostatic properties and signal generation. Manganese dioxide ($MnO_2$), with high redox activity, enables moisture and oxygen sensing, critical for modified atmosphere packaging (Nikolić et al., 2021; Nami et al., 2024).

Nanotechnology is rapidly transforming the food industry, with nanomaterials offering compelling advantages in efficiency, sustainability, and product quality. These innovations align with several United Nations Sustainable Development Goals (SDGs), particularly those related to health, food security, and sustainability. Applications such as nano-enabled nutritional supplements and smart packaging with antimicrobial or freshness-sensing functions enhance food quality, safety, and shelf life, supporting SDG 2 (Zero Hunger), SDG 3 (Good Health and Well-being), and SDG 12 (Responsible Consumption and Production). To a lesser extent, they may contribute to SDG 13 (Climate Action) by reducing food waste and associated emissions. Together, these benefits could tip the scales toward the widespread adoption of nanomaterials in food applications. However, to ensure these benefits do not come at the cost of unintended health risks, it is crucial to carefully assess their safety. Striking the right balance in the risk-benefit ratio of nanomaterials requires a comprehensive evaluation of their potential toxicity, particularly their interactions with biological systems.

### 1.2. Toxicological profile and Toxicity of Food-Related Nanoparticles

A key downside of the increasing use of engineered nanomaterials in food is growing concern over their potential toxicity. Once ingested, NPs undergo complex transformations in the gastrointestinal (GI) tract that influence their absorption, metabolism, and toxicity. These effects depend largely on physicochemical properties such as size, charge, solubility, and composition. This section reviews current literature on the toxicological profiles and health risks of food-related nanomaterials.

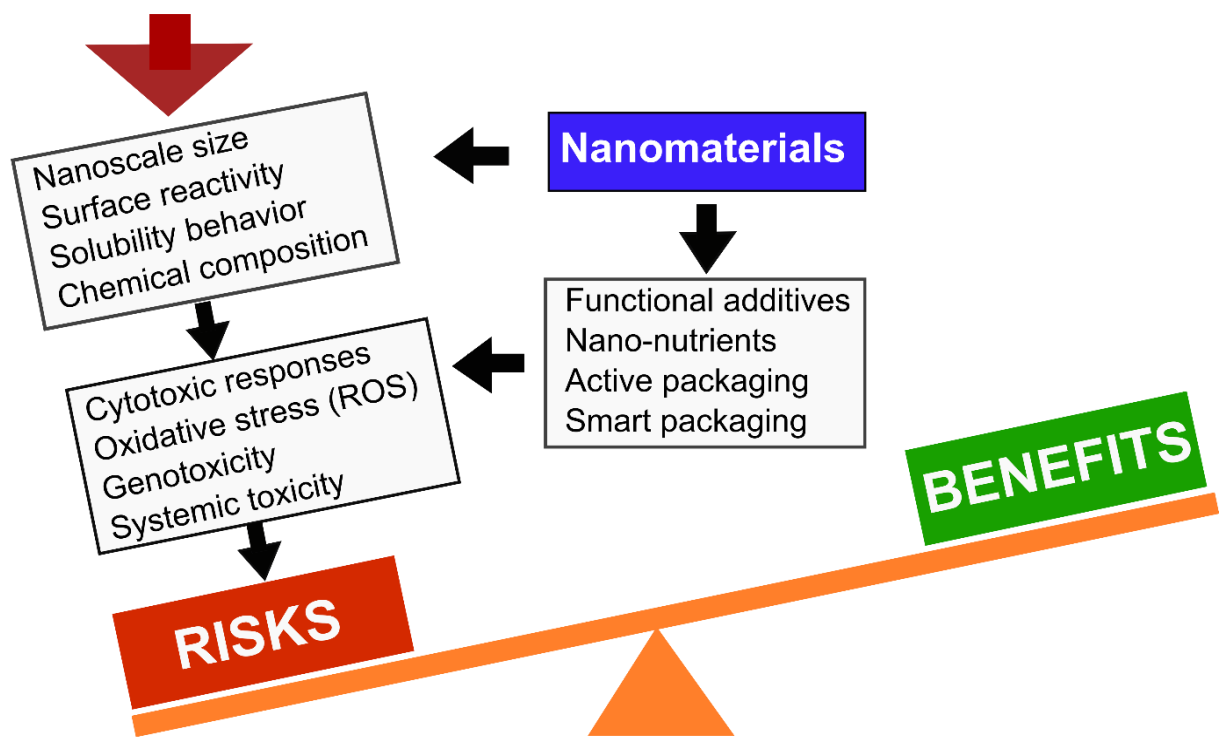

*Figure 2.* Schematic balance highlighting the risks of nanomaterials in food-related applications.

Size-dependent toxicity plays a pivotal role in NP interactions with biological systems. Smaller NPs (<100 nm) possess higher surface-area-to-volume ratios, which increase their reactivity and potential to induce oxidative stress. This relationship between size and toxicity is well documented in several comprehensive reviews. For example, Card et al. (2011) emphasize that physicochemical properties such as particle size and surface area are critical determinants of NP behavior in biological systems, influencing their deposition, distribution, and clearance. McCracken et al. (2016) provide further insights through their bibliographic analysis. They specifically highlight that smaller particles, particularly nanoscale silica ($SiO_2$), $TiO_2$, ZnO, and Ag, are more likely to induce cytotoxic effects due to their capacity to generate oxidative stress and inflammation in gastrointestinal epithelial cells. Adding to these concerns, Wang et al. (2021) draw attention to the fact that ultrafine NPs, particularly those smaller than 50 nm, are capable of crossing epithelial barriers. Once beyond these barriers, they can enter systemic circulation and accumulate in vital organs such as the liver, kidneys, and brain. This ability to bypass traditional detoxification mechanisms raises concerns about their potential to provoke chronic inflammation and oxidative damage. Similar observations are made by Jain et al. (2018), who report that nanoscale food additives, including $TiO_2$ and $SiO_2$, display size-dependent cytotoxicity. Their review indicates that smaller particles elicit stronger inflammatory responses in gut epithelial cells compared to their larger counterparts. Expanding on these findings, Zickgraf et al. (2023), in a meta-analysis of recent studies, confirm that NPs below 100 nm can significantly alter gut microbiota composition. Such disruptions are associated with reduced production of beneficial short-chain fatty acids (SCFAs), potentially compromising immune homeostasis and contributing to systemic effects. Further corroborating these impacts, Xie et al. (2022) demonstrate that exposure to nanomaterials, particularly Ag,

TiO$_2$, ZnO, and SiO$_2$, can modulate gut microbiota, often leading to dysbiosis and metabolic disruptions. Similarly, Jiang et al. (2024a) emphasize that inorganic NPs like TiO$_2$ and SiO$_2$ not only penetrate intestinal barriers but can also induce chronic toxicity and disrupt intestinal homeostasis.

However, size is not the sole determinant of NP toxicity. Surface charge and functionalization also play significant roles in shaping their toxicological profiles. For instance, positively charged (cationic) NPs are frequently more cytotoxic due to their strong electrostatic attraction to negatively charged cell membranes (Jain et al., 2018). This interaction facilitates cellular uptake but can also cause membrane disruption and subsequent cytotoxic effects. Card et al. (2011) noted that gold nanorods capped with cationic cetyltrimethylammonium bromide kill ~70 % of enterocyte-like cells; adding a polyacrylic acid or poly(allylamine) over-coating reduces their toxicity about fivefold, underscoring the decisive role of surface chemistry. In line with this, Zickgraf et al. (2023) and McCracken et al. (2016) also noted that surface functionalization can profoundly influence NP interactions with microbial populations in the gut, potentially affecting nutrient absorption and metabolic processes.

In addition to surface charge, solubility and biodegradability are key factors influencing NP toxicity. For instance, poorly soluble metal oxide NPs like TiO2 NPs tend to accumulate in tissues over time, increasing the risk of bioaccumulation and chronic toxicity whereas polystyrene NPs are absorbed only to a limited extent (Card et al., 2011). McCracken et al. (2016) also noted that amorphous SiO2 tends to be more soluble and is generally considered safer than its crystalline counterpart whose lower solubility is associated with heightened cytotoxicity. Zickgraf et al. (2023) also reported that poorly soluble TiO2 NPs can alter gut microbiota composition and provoke inflammatory responses. Supporting these concerns, Mcclements and Xiao (2017) emphasized that even minimal absorption of Ag NPs can result in their bioaccumulation in critical organs, such as the liver, spleen, and kidney. Additionally, Gupta et al. (2024) examine the role of nanomaterials used in food packaging. Their review focuses on the migration of NPs specifically Ag, TiO$_2$, and ZnO, from packaging into food products. This migration raises further concerns about chronic human exposure and the potential for long-term bioaccumulation.

The chemical composition of NPs also critically influences their biological effects. Card et al. (2011) highlight that metal-based NPs like Ag and ZnO release toxic ions that can induce oxidative stress, mitochondrial damage, and DNA fragmentation. Echoing these findings, McCracken et al. (2016) emphasize that Ag and ZnO NPs promote oxidative and inflammatory pathways. Wang et al. (2021) further report that TiO$_2$, despite its widespread use as a food additive, has been linked to gut inflammation and potential carcinogenicity. These findings have contributed to regulatory actions, such as the EFSA's decision to ban TiO$_2$ as a food additive. Jiang et al. (2024a) corroborate these concerns, noting that TiO$_2$ and SiO$_2$ NPs can compromise intestinal barrier integrity and induce chronic toxicity. Interestingly, their review also acknowledges potential beneficial roles for certain food-derived NPs in enhancing barrier function and improving nutrient delivery, illustrating the complexity of NP interactions within biological systems. Finally, Li et al. (2024) provide additional evidence that food additives, including inorganic NPs like TiO$_2$ and Ag, can disrupt gut microbiota and compromise intestinal barrier function. Their findings link these disruptions to broader health concerns, including metabolic disorders and neurobehavioral effects. The table below presents selected quantitative toxicological data for well-characterized nanoparticles, compiled from representative case studies. It includes key parameters such as $IC_{50}$ concentrations, exposure durations, nanoparticle size and surface charge, as well as the main mechanistic pathways involved in their toxicological effects.

**Table 1. Quantitative nanotoxicology data for representative nanoparticles used in food products**

| NP type | Size/ Charge | Exposure duration | IC50 | Mechanistic effect | Ref |
|---|---|---|---|---|---|
| $TiO_2$ | $TiO_2$-NPs 60 nm (rutile) and $TiO_2$ E171 (anatase)/not reported | 72 h (In vitro Colon Cancer Cells) | $TiO_2$-NPs 60 nm: 41.1 mg/L; $TiO_2$ E171: 3.45 mg/L | Genotoxic effect induced by moderate DNA damage. E171: cell growth inhibition and higher level of cytotoxicity and genotoxicity | Ferrante et al., 2022 |
| ZnO (in additive solvents) | ~78 nm/+20 to +30 | 24 h (In vitro Colon Cancer Cells) | ZnO:170 µg/mL, 177 µg/mL in methanol, 95 µg/mL in glycerin; 80 µg/mL in propylene glycol | Cell proliferation inhibition, membrane damage and ROS generation, higher in glycerin and propylene glycol. | Lee et al., 2023 |
| $Fe_3O_4$ | ~11 nm/ -12.20 mV | 24–48 h (In vitro endothelial cells) | 79.13 µg/mL | Cytotoxicity, oxidative stress, ROS production, and mitochondrial membrane dysfunction | Siddiqui et al., 2023 |
| $SiO_2$ | ~ 55-60nm/ -30.33±1.35 mV, | 24–48 h (In vitro: endothelial cells) | 50 µg/mL | Oxidative stress induced by ROS generation, apoptosis via Bax/Bcl-2 imbalance, mitochondrial dysfunction | Guo et al., 2016 |

Despite a growing body of research on the toxicological effects of food-related nanoparticles (NPs), significant inconsistencies persist, hindering clear conclusions about their long-term safety. Reviews by Card et al. (2011) and McCracken et al. (2016) point to core issues such as inconsistent NP characterization, short study durations, and the lack of standardized testing protocols. These methodological limitations undermine the reliability and reproducibility of toxicity assessments, a concern echoed in recent meta-reviews and regulatory evaluations.

Siivola et al. (2022) conducted a systematic quality review of nanomaterial genotoxicity studies and found that many published results fail to meet regulatory standards due to inadequate physico-chemical characterization, non-validated genotoxicity assays, and omission of critical exposure parameters such as solubility, agglomeration, and surface modifications. This aligns with observations in Yamani et al. (2024), which reported that over 90% of nanotoxicity studies published before 2014 neglected to account for NP-induced assay interference, leading to potentially spurious outcomes.

One prominent area of uncertainty involves the interactions between NPs and the gut microbiome, as well as their potential for chronic toxicity. Studies report conflicting outcomes, ranging from substantial microbial disruption and metabolic changes to minimal or no observable effects (Zickgraf et al., 2023; McClements and Xiao, 2017; Wang et al., 2021). These discrepancies are frequently attributed to particle-dependent properties such as size, surface charge, and solubility, along with inconsistent use of biological models. Møller et al. (2024) highlighted this issue explicitly, demonstrating how oral exposure to particles such as $TiO_2$ and AgNPs yields heterogeneous genotoxicity outcomes, sometimes positive, often null, depending on the assay used, lab practices, and animal model. Their meta-analysis of DNA strand break data from comet assays revealed substantial between-study variability, further emphasizing poor reproducibility even within a specific particle class.

These insights highlight the urgent need for more complex and physiologically relevant test models. A major limitation across many current studies is the use of overly simplistic in vitro systems that fail to mimic the human gastrointestinal tract. Halappanavar et al. (2021) discussed how traditional in vitro models often lack essential components like mucus layers, immune cells, and dynamic exposure mechanisms and stressed that assay endpoints in the literature remain poorly matched across platforms.

Without robust data and clear risk assessments, the risk–benefit scale tilts unfavorably, weighing more heavily on potential harms and hindering the adoption of nano-enabled food products. To address this imbalance, emerging technologies such as MPS, which replicate human tissue architecture, dynamic flow, and multi-organ interactions, can enhance the accuracy of safety assessments and help bridge the gap between in-vitro data and real-world outcomes (Ashammakhi et al., 2020; Lu and Radisic, 2021; Xavier et al., 2021). By reducing scientific uncertainty and more closely replicating human gut conditions, these models can help rebalance the scale. As risks become more measurable and manageable, the weight may begin to shift toward the benefits, supporting the careful and responsible integration of nanotechnology in food systems, where innovation is guided by evidence-based safety and public trust.

2. **Advancing Nanotoxicity Assessment with Microphysiological Systems**

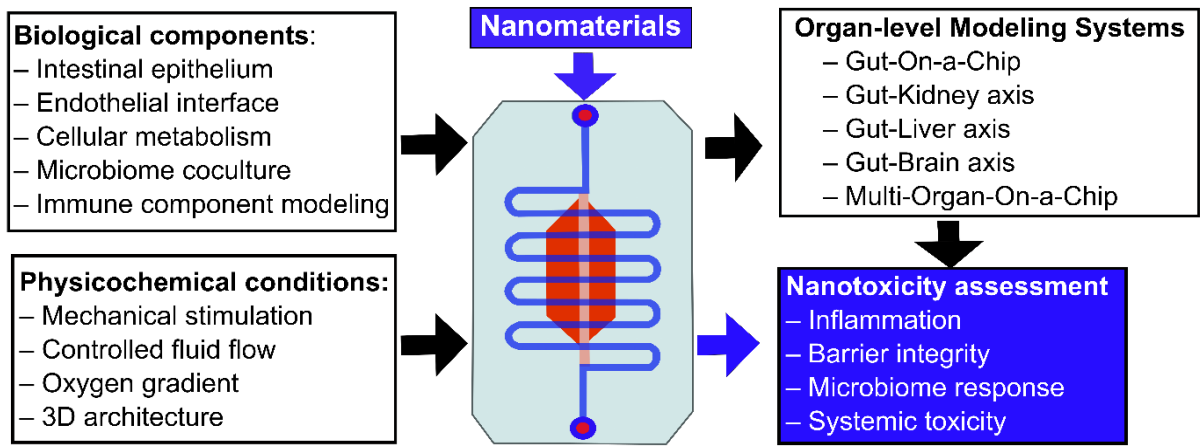

*Figure 3. Schematic overview of microphysiological system (MPS) integration for nanotoxicity assessment*

Recent advances in GOC technologies have significantly improved the ability to replicate the complex environment of the human gastrointestinal (GI) tract. Innovations in microfluidics, biomaterials, and tissue engineering now enable the precise simulation of intestinal dynamics, oxygen gradients, and epithelial barrier architecture (Figure 3), highlighting the advantages of MPS over conventional in vitro and in vivo models (Guo et al., 2023, Avula et al., 2024). To illustrate these advantages, a comparative overview is presented in Table 2. In light of these insights, the following section provides a structured, multi-level analysis of how MPS can model both the gut and its interactions with other organs (the gut–organ axis), not only in general physiological contexts but also in nanotoxicological assessments.

**TABLE 2: Comparative Overview of MPS, conventional In Vitro, and Animal Models.**

| Indicator | MPS | 2D In Vitro Models | Animal Models |
|---|---|---|---|
| **Biological relevance** | ✓ **Accessible** (mucus, microbiota and immune components) | ⚠ **Limited** (lack mucus, microbiota, immune components; static exposure) | ⚠ **Variable** (whole-body ADME; inter-species differences) |
| **Inter-organ communication** | ✓ **Possible** in multi-organ chips | ✗ Absent | ✓ **Present** (systemic) |
| **Dynamic physiological environment** | ✓ Simulates flow, peristalsis, gradients | ✗ Static | ✓ **Present** (though not controlled) |
| **Mechanistic insight** | ✓ **Accessible** (Real-time TEER, cytokine/ROS sensors, live imaging) | ⚠ **Limited** (cytotoxicity / barrier endpoints) | ⚠ **Limited** (Good systemic readouts) |

| Throughput | ⚠ Emerging | ✓ Established (96/384-well formats) | ✗ Low |
|---|---|---|---|

## 2.1. Simulating Gut-On-a-Chip Physicochemical Features

Dynamic fluid flow is a fundamental feature in GOC, as it closely replicates key physiological conditions of the human gastrointestinal tract. By simulating shear stress, promoting continuous nutrient delivery, and enabling efficient waste removal, dynamic flow plays a vital role in sustaining epithelial cell viability, promoting differentiation, and maintaining functional barrier integrity. A seminal advancement in this field was introduced by Kim et al. (2012), who developed one of the earliest GOC models that integrated peristalsis-like mechanical strain with laminar microfluidic flow. Their system utilized microchannels lined with human intestinal epithelial cells exposed simultaneously to cyclic mechanical forces and controlled fluid movement. This dual-stimulus approach facilitated microbial co-culture, enhanced epithelial polarization, and led to the spontaneous formation of villus-like structures, marking a key innovation in biomimetic modeling of the gut.

Building on this foundation, de Haan et al. (2021) proposed a compartmentalized GOC platform that merged digestion simulation with compound absorption profiling. Their modular system emulated sequential digestive phases, oral, gastric, and intestinal, through enzyme-laden microreactors. Downstream, digested samples were perfused over an intestinal epithelial barrier under continuous flow, enabling real-time mass spectrometric analysis. This integration significantly enhanced the physiological relevance of pharmacokinetic assessments and allowed precise measurement of compound bioavailability.

Further advancing this paradigm, Xavier et al. (2023) designed a two-module microfluidic platform that comprehensively replicates the gastrointestinal journey, from ingestion to absorption. Their system comprises a Digestion-Chip, simulating enzymatic breakdown across oral, gastric, and intestinal phases, coupled with a Gut-Chip, which hosts Caco-2/HT29-MTX co-cultures under perfusion. Notably, the Gut-Chip supported rapid epithelial differentiation and the emergence of 3D villi-like morphology, closely mirroring in vivo intestinal architecture. Adding a molecular perspective, Kulthong et al. (2021) performed transcriptomic profiling that underscored the biological importance of dynamic flow. Their study revealed that Caco-2 cells cultured under flow conditions within GOC devices exhibit gene expression profiles more closely aligned with native human intestinal tissue compared to static culture systems. These findings underscore the critical role of fluid dynamics not only in structural and functional epithelial maturation but also in achieving molecular fidelity in vitro. Beyond fluid flow, mechanical stimulation, particularly peristalsis-like strain, has proven vital for reproducing the dynamic nature of intestinal motility. Kim et al. (2012) were among the first to apply cyclic mechanical strain to intestinal epithelium in a microfluidic environment, mimicking the effects of gut peristalsis and enhancing structural and functional maturation of the epithelial layer. Further exploring this dimension, Gérémie et al. (2022) investigated the effects of cyclic stretching on Caco-2 monolayers. Their findings demonstrated that prolonged mechanical strain leads to epithelial reorganization and alignment, emphasizing the importance of mechanical forces in epithelial polarity and integrity. This mechanotransduction appears to be a key regulator of gut epithelium morphology. Apostolou et al. (2021) built on these insights by developing a colon Intestine-Chip featuring co-cultured organoid-derived epithelial cells and endothelial cells under flow and mechanical strain. Their system allowed investigation of inflammatory cytokine responses, tight junction remodeling, and barrier dysfunction in a controlled microenvironment that closely mimics in vivo conditions. The integration of mechanical and biological stimuli in this model makes it especially useful for studying intestinal permeability and inflammation. The intestinal environment exhibits steep radial oxygen gradients, with oxygenated tissue coexisting alongside anaerobic microbiota. Accurately recreating this dual environment is crucial for studying host-microbe interactions. Several models have addressed this challenge through creative engineering of oxygen control mechanisms. Jalili-Firoozinezhad et al. (2019) developed an anaerobic intestine-on-a-chip system capable of supporting stable co-culture of complex gut microbiota and human intestinal epithelium. Their device included microscale oxygen sensors and isolated luminal compartments, maintaining physiologically relevant oxygen gradients and enabling sustained microbial diversity. Liu et al. (2023) proposed a complementary strategy using simulation-guided chip design to create controllable oxygen gradients. Their system facilitated co-culture of Bifidobacterium bifidum and epithelial cells, supporting the study of microbial influence on inflammatory bowel disease. This model ensured anaerobic conditions in the lumen while maintaining oxygenation of the basal compartment.

Advances in MPS have enabled sophisticated GOC platforms, offering a strong foundation for biological modeling. However, fully realizing their potential requires integration with a deeper biological understanding of the intestinal barrier. While features such as fluid flow, mechanical strain, and oxygen gradients enhance physiological relevance, replicating the gut's structural and functional complexity, particularly the mucus–epithelium interface, demands equally advanced biological modeling. The next frontier lies in mimicking native barrier architecture, including crypt-villus structures, stem cell-driven regeneration, and the mucus layer's dual roles in protection and absorption.

### 2.2. The Intestinal Barrier: From Structure to Integrity Assessment

The intestinal barrier serves as a critical gatekeeper, regulating what enters the bloodstream and what is kept out. Composed of a single epithelial layer sealed by tight junctions and protected by a mucus layer, it facilitates nutrient absorption while blocking harmful agents. However, ingested engineered nanomaterials can compromise this defense. However, ingested engineered nanomaterials can compromise this defense. TiO2NPs, for example, have shown to disrupt tight junction integrity, thereby increasing paracellular permeability and provoking oxidative stress–driven epithelial dysfunction. These disturbances coincide with elevated pro-inflammatory cytokines and shifts in gut-microbiota composition, collectively impairing intestinal homeostasis (Schwarzfischer & Rogler, 2022). These insights highlight the need for advanced in vitro models that accurately replicate key features of the gut barrier, including the crypt–villus architecture and the mucus layer's dual role in protection and absorption.

One of the earliest dynamic systems, developed by De Gregorio et al. (2019), integrated a 3D human intestinal equivalent composed of Caco-2 epithelial cells layered over a myofibroblast-derived extracellular matrix (ECM) under flow. This configuration promoted epithelial polarization, tight junction formation, and mucus secretion. To explicitly model the mucus barrier microstructure, Sardelli et al. (2024) introduced a microfluidic system incorporating a 3D bioinspired mucus gel (I-Bac3Gel) with viscoelastic and microstructural features resembling native colonic mucus. While powerful for studying lipopolysaccharides diffusion and bacterial survival, the system lacked an epithelial layer, precluding epithelial–mucus interaction studies. Addressing this, Lee et al. (2023) designed a gut-mucus chip with mucin-coated Caco-2 monolayers under dynamic flow. Although it introduced cellular elements, its reliance on exogenous mucus and a single epithelial lineage restricted physiological complexity.

Recent developments have emphasized multicellularity and biomimetic architecture. Moerkens et al. (2024) used iPSC-derived intestinal organoids in a dual-channel chip to recapitulate absorptive, secretory, mesenchymal, and neural cell populations. The model self-organized into villus-like structures and established barrier function under flow and growth factor gradients, though the system remained technically demanding and time-intensive. In parallel, Almalla et al. (2024) combined decellularized ECM-based 3D-printed scaffolds with primary and immortalized cells under flow. This approach promoted spontaneous differentiation into mucus-secreting lineages, offering an elegant integration of mechanical and biochemical cues, albeit without real-time functional readouts.

Valibeknejad et al. (2025) presented a complementary approach focusing on the biophysical properties of the mucus barrier. Their biosimilar mucus-on-chip platform employed tunable hydrogel formulations to simulate healthy and diseased states. Using validated computational fluid dynamics simulations, they quantified mucus dislodgement and particle diffusion under shear flow. Although epithelial-free, this system offered predictive insights into mucus mechanics, reinforcing the value of coupling experimental models with in silico tools.

A major leap forward came with Palamà et al. (2025), who introduced a PDMS (Polydimethylsiloxane)-free, double-flow GOC platform. By co-culturing Caco-2 and HT-29 cells under apical and basal flow, they achieved rapid epithelial maturation, tunable mucus production, and tight junction formation within 7–10 days. Crucially, they developed a computational TEER (transepithelial electrical resistance) model to predict barrier integrity from cell composition, providing a bridge between structural features and quantitative function. The work by Palamà et al. (2025) exemplifies a critical transition in GOC research from static morphological characterization to predictive, functional modeling of intestinal barrier performance through real-time, quantitative evaluation. Central to this evolution is TEER, a well-established technique increasingly embedded into organ-on-chip platforms to provide high-resolution, dynamic readouts of epithelial integrity under flow.

Giampetruzzi et al. (2022) developed a transparent-electrode-based system with a microbubble-tolerant algorithm, enabling continuous TEER monitoring under flow. Their impedance spectroscopy approach allowed robust tracking of epithelial maturation over 24 days. Advancing further, Lucchetti et al. (2024a) embedded multiple flexible thin-film electrodes into the human microbial crosstalk GOC system, enabling spatially resolved, real-time TEER measurements. Their finite element modeling corrected for geometric distortions in current flow, revealing heterogeneity in barrier development and recovery.

Taking a biofabrication approach, Vera et al. (2024) created a 3D bioprinted hydrogel GOC that combined Caco-2 epithelial cells with stromal fibroblasts in a villus-mimetic matrix. Integrated platinized electrodes enabled TEER monitoring under flow, capturing barrier formation dynamics with physiological accuracy. This modular system was especially well-suited for probing epithelial–stromal interactions in real time.

Other studies applied TEER in more specialized contexts. Brandauer et al. (2025) modeled senescence-associated barrier dysfunction by exposing Caco-2 cells to doxorubicin. Despite increased claudin-2 expression typically associated with leakiness, they observed TEER elevation, attributed to hypertrophic cell morphologies. Their membrane-integrated TEER sensors provided continuous monitoring of age-related barrier alterations.

Furthermore, Morelli et al. (2025) employed the OrganoPlate® and OrganoTEER® systems to study Candida albicans. Exposure to candidalysin, a fungal cytolytic toxin, led to TEER loss, increased epithelial permeability, and inflammatory cytokine release. Their findings, validated in patient-derived colon organoids, underscore TEER's value in toxicological and pathogen-host interaction studies.

These systems reflect a rapid shift from static, monotypic cultures to dynamic, multicellular, and physiologically relevant GOC models. TEER integration has become central, enabling real-time, quantitative monitoring of barrier function in increasingly complex platforms. Recent advances in epithelial modeling on chip now allow for more accurate replication of intestinal mechanics and function. Yet, epithelial integrity is only one aspect of gut physiology. To better emulate the human gut, the next step is integrating microbial communities and immune components. Coupling barrier models with host–microbiome and immune system interactions is essential for advancing nanotoxicity assessment within a more holistic GOC paradigm.

## 2.3. Enhancing Gut Microphysiological Systems: Integrating Microbiota and Immune components

Along the gastrointestinal lining, the gut microbiota forms a dynamic and diverse ecosystem of bacteria, fungi, and other microorganisms that play vital roles in digestion, metabolism, immune regulation, and even mental health. The development of GOC platforms now offers a powerful tool to model these host–microbiota interactions in physiologically relevant conditions.

A foundational study by Kim et al. (2016) introduced a microfluidic GOC system designed to simulate the mechanical and structural environment of the human intestine. By applying fluid shear and peristalsis-like deformations to a Caco-2-derived villus epithelium, the platform enabled long-term co-culture with human gut microbiota. Microbial colonies established between villi, maintaining viability alongside differentiated epithelial cells for days to weeks. However, while this model marked a significant advancement in reproducing dynamic mechanical cues and epithelial topology, it lacked control over oxygen gradients, limiting its compatibility with obligate anaerobes.

Focusing on epithelial maturation and barrier functionality, Jeon et al. (2022) introduced a perfused microfluidic model capable of promoting polarization, mucus production, and tight junction formation. The incorporation of TEER electrodes enabled real-time monitoring of epithelial integrity in response to probiotic strains. While this system demonstrated enhanced epithelial differentiation compared to the models developed by Kim et al. (2016), it still lacked controlled oxygen gradients necessary to support obligate anaerobes and did not incorporate biomechanical stimulation. To overcome oxygenation constraints, Shin et al. (2019) developed an anoxic–oxic interface-on-a-chip by engineering a dual-channel microfluidic device with opposing oxygen supplies, successfully supporting the growth of strictly anaerobic bacteria such as Bacteroides fragilis. The model maintained epithelial barrier function, monitored via TEER, under dynamic co-culture conditions. This improved microbial physiological fidelity compared to Kim et al. (2016)'s system but still relied on Caco-2 cells and lacked additional layers of tissue complexity, including immune and endothelial components.

Recently, Kim et al. (2024a) introduced an in vitro colon model that mimics the human colonic environment by integrating a functional mucus layer and an anaerobic air-liquid interface. The system maintains an oxygen-depleted apical side and supports coculture with both facultative and obligate anaerobic gut bacteria. The mucus layer provides partial bacterial exclusion, indicating protective properties. This dual-feature model addresses key limitations in earlier systems, though further optimization is needed for long-term viability and complex microbial communities.

Wang et al. (2024a) enhanced disease relevance by constructing a GOC model colonized with microbiota derived from patients with depression. This platform recapitulated hallmarks of intestinal dysfunction observed in depressive disorders, such as reduced barrier integrity, increased inflammatory cytokines, and altered neurotransmitter metabolism. This approach represented a meaningful step toward personalized modeling. In terms of scalability and real-time visualization, Lee et al. (2024) presented a Gut Microbiome-on-a-Chip platform featuring a 3D stratified epithelium cultured under dynamic fluidic conditions. The system enabled direct observation of microbial colonization, carcinogenic pathogenesis, and probiotic effects, providing a powerful platform for high-throughput screening. Compared to earlier systems, Lee et al. (2024)'s chip enhanced imaging and throughput while maintaining a focus on epithelial–microbiota interactions. However, like previous models, it lacked immune component integration, limiting its ability to model chronic inflammation and full gut immunophysiology.

The gut immune system is shaped by continuous interactions with the intestinal microbiota, mediated by microbial metabolites, pattern recognition receptors, and cytokine signaling. Among the key modulators are short-chain fatty acids (SCFAs), such as butyrate and propionate, which promote regulatory T cell (Treg) activation, enhance tight junction integrity, and suppress inflammation via G-protein-coupled receptors and histone deacetylase (HDAC) inhibition. Microbial components also engage receptors like Toll-like receptors (TLRs), fostering immune tolerance and epithelial defense. Additionally, commensals modulate cytokine profiles by promoting anti-inflammatory mediators (e.g., IL-10, IL-22) while suppressing pro-inflammatory ones (e.g., TNF-α, IL-17) (Lamas et al., 2020; Zickgraf et al., 2023). This intricate crosstalk is especially relevant in the context of nanotoxicology, where ingested nanomaterials may disrupt microbiota composition or immune signaling, altering host immune homeostasis, exacerbating inflammation, and increasing systemic toxicity (Zickgraf et al., 2023). These microbiota–immune interactions are therefore critical to consider when modeling gut responses in vitro for the safety assessment of nano-enabled food products.

Initial advancements in immune integration within GOC models were led by Shin et al. (2018) and Maurer et al. (2019), enabling the study of host-microbiota-immune interactions under physiologically relevant conditions. Shin et al. (2018) developed a dynamic platform that mimicked peristalsis-like motions, fluid flow, and oxygen gradients, allowing co-culture with anaerobic bacteria and integration of peripheral blood mononuclear cells (PBMCs). This setup enabled the dissection of how epithelial barrier dysfunction, in combination with microbial signals and immune presence, triggers oxidative stress and proinflammatory cytokine release. Maurer et al. (2019) extended this concept by constructing a 3D immunocompetent GOC model incorporating epithelial, endothelial, and tissue-resident innate immune cells (macrophages and dendritic cells) under dynamic perfusion. The system supported microbial colonization, cytokine profiling, and inflammation modeling, demonstrating both high structural complexity and robust immunological relevance. While Shin et al. (2018)'s model offers strong control over microbial dynamics and barrier function, Maurer et al. (2019)'s platform provides a more integrated immune architecture, enabling deeper insights into mucosal immunity and host–pathogen interactions. Moving further in immune component integration, De Gregorio et al. (2022) developed an immunoresponsive microbiota-GOC combining epithelial and stromal cells with PBMCs, along with a custom anaerobic chamber that supports stratified colonization of strict anaerobes across a luminal-serosal oxygen gradient. This configuration enabled detailed analysis of ROS generation, cytokine secretion, and stromal remodeling under inflammatory conditions. Nonetheless, long-term stability and full adaptive immune function were not fully achieved.

Humayun et al. (2022) complemented this direction by creating a 3D MPS to investigate innate immune responses specifically neutrophil and natural killer cell activity, against Toxoplasma gondii, offering mechanistic insights into early host defense but lacking microbiota and tissue oxygen levels.

Translating these features to personalized immunology, Ballerini et al. (2025) developed a GOC model seeded with human intestinal organoids, microvascular endothelial cells, and patient-derived fecal microbiota. This platform predicted clinical responses to immune checkpoint inhibitors in melanoma patients, underscoring the microbiome's role in immunotherapy outcomes. However, the lack of integrated immune cells limited insights into immune cell–microbe signaling pathways.

Among the most advanced systems to date, Zhang et al. (2024) introduced a dynamic GOC model co-culturing colonic epithelial cells, strict anaerobic microbiota, and both innate and adaptive immune cells. Featuring fluidic oxygen gradient control, long-term co-culture stability, and real-time TEER monitoring, this platform captured chronic inflammatory responses, immune activation, and microbial translocation, establishing a high-fidelity model of host-microbiota-immune interactions.

State-of-the-art GOC models have transformed intestinal research by enabling the study of gut inflammation, microbiome interactions, and immune responses. By replicating key features, such as peristalsis, oxygen gradients, stratified epithelium, and microbial colonization, they offer powerful tools for assessing the safety of nano-based food additives and packaging. For instance, models that incorporate anaerobic microbiota and immune cells provide critical insights into host–microbe–immune crosstalk, which likely influences NP reactivity and intestinal immune responses.

By replicating features like peristalsis, oxygen gradients, and microbial colonization while incorporating immune cells and anaerobic microbiota to capture host-microbe-immune crosstalk, GOC models offer powerful tools for assessing the safety of nano-based food products.

The table below summarizes selected state-of-the-art GOC models by year of publication, highlighting core features and physiological relevance across varying levels of complexity. This comparison illustrates the technological progression toward more advanced platforms for studying host-microbiome interactions and toxicology.

**Table 3: Comparative Overview of Gut-on-Chip Models.**

| Study | Model Description | Main Features | Physiological Relevance |
|---|---|---|---|
| Kim et al., 2016 | GOC with multiple commensal microbes coculture | Shear flow + cyclic strain, microbial co-culture, TEER measurements confirm barrier maturation. | High (intestinal villi lined by 4 epithelial cell types, dynamic interface) |
| Shin et al., 2018 | GOC with co-culture with gut microbes and immune cells | Shear flow + mechanical motions, co-culture with microbiota; cytokine response + barrier integrity TEER monitoring) | High (fluid flow + Peristalsis-like flow, inflammatory host–microbiome cross-talk) |
| Maurer et al., 2019 | 3D immunocompetent intestine-on-chip | Shear flow, Candida/Lactobacilli interactions, mucosal immunity, epithelial-endothelial barrier. | High ($O_2$ control, tissue complexity: epithelium-endothelium microbiota + immune co-culture) |
| Shin et al., 2019 | Gut microbiome + intestinal epithelium in an anoxic-oxic Interface-on-a-Chip | Shear flow, cyclic strain, host-gut microbiome interactions, immune cell response, barrier integrity via TEER recording | Moderate to High ($O_2$ gradient + peristalsis, microbiome interactions + epithelium interactions) |
| Jeon et al., 2022 | Three-channel microfluidic-based GOC | Inflammation modeling, probiotic co-culture under low shear flow, epithelial and endothelial cells + TEER recording | Moderate (epithelium-endothelium + bacterial cells, mechanical flow) |
| Humayun et al., 2022 | Intestinal MPS with 3D tubular structure of epithelium | Host-parasite innate immune response. | High (epithelium-endothelium + immune) |
| De Gregorio et al., 2022 | Microbiome-GOC with 3D epithelium | Shear flow, stromal reshaping, ROS, cytokines, TEER used to monitor barrier dysfunction. | Very High (tissue complexity + inflammation, + $O_2$ gradient) |
| Kim et al., 2024a | MPS Anaerobic ALI + mucus + host-colon bacteria | Commensal anaerobe-host interactions, in vivo colon's hypoxic condition. | High (mucus + $O_2$ gradient + obligate anaerobic gut bacteria) |
| Lee et al., 2024 | Stratified 3D gut + microbial visualization (e.g., ETBF, Lactobacillus) | Shear flow, microbial competition, validated intestinal epithelium with villi structure. | High (spatial resolution + $O_2$ gradient, gut-microbial interfaces + co-culture epithelial cells) |
| Wang et al., 2024a | Microbiota + gut barrier on-a-chip | Shear flow, depression biomarkers, anaerobic bacteria, 3D villi structure. | High (disease-specific microbial profile, $O_2$ gradient) |
| Zhang et al., 2024 | Immune-competent gut MPS + colonic epithelium | Shear flow, immune modulation by commensals, tight junction integrity confirmed via TEER | Very High (immune + anaerobe bacteria + $O_2$ gradient) |
| Ballerini et al., 2025 | GOC with iPSC-organoids + human feces | Shear flow + peristaltic-like motion, Predictive of immune checkpoint inhibitor response (melanoma) | Very High (epithelial, microvascular, immune and microbiome components) |

Physiological relevance is categorized as:
- Very High: ≥5 features; advanced physiological mimicry; includes both immune cells and microbiota (including anaerobes).
- High: 3–4 features; substantial mimicry; includes either immune cells or microbiota (or both in partial co-culture).
- Moderate: 1–2 features; basic physiological mimicry; lacks integrated immune or microbial components.

Despite these advances, current GOC systems remain constrained by several limitations. Standardized protocols for NP dosing and chronic exposure modeling are still lacking, and maintaining long-term co-culture of microbiota and immune cells remains technically challenging. These gaps hinder accurate assessment of cumulative or subtle toxic effects over extended timescales. More fundamentally, GOC platforms focus on localized intestinal events and do not account for systemic physiological processes. Key determinants of nanomaterial toxicity such as first-pass metabolism, immune cell migration, neuroendocrine signaling, and multi-organ distribution, cannot be captured in isolated gut models. Importantly, the absorption, biotransformation, and distribution of NPs involve interactions across multiple organs, including the liver, kidney, vasculature, and central nervous system. These complexities underscore the need for integrated models reflecting the physiological connectivity of the human body. To meet this challenge, multi-organ-on-a-chip (MOC) platforms have emerged as a promising extension of single-organ systems. By linking gut models with other tissues via microfluidic circuits, MOC technologies can enable real-time tracking of NP absorption, circulation, metabolism, and downstream toxicological effects. The following section explores current progress in MOC systems.

### 2.4. Multi-Organ Integration: Capturing Systemic Nanotoxicity

Once ingested, nano-enabled food products can cross the intestinal barrier, enter systemic circulation, and accumulate in various organs, raising concerns about their potential to induce systemic toxicity. This section presents state-of-the-art research in MOC technologies, with a particular focus on recent advancements in modeling interconnected gut-organ systems.

As the body's central detoxification hub, the liver is among the first organs exposed to nanoparticles absorbed through the gut, making the gut-liver axis a critical model for studying first-pass effects. A notable example is the study by Wang et al. (2024b), who developed a gut–liver-on-a-chip incorporating biomimetic intestinal peristalsis and dynamic hepatic flow to closely replicate in vivo physiology for evaluating microplastic toxicity. To this end, they perfused fluorescently labeled polystyrene microplastics through the system at varying concentrations after intestinal barrier maturation. TEER measurements and fluorescence imaging were used to assess barrier integrity and microplastic transport, while downstream hepatic uptake was examined via immunostaining. Cellular analyses revealed dose-dependent liver toxicity, characterized by ROS, elevated inflammatory cytokine (TNF-α) levels, and reduced cell viability. Notably, simulated peristalsis enhanced intestinal barrier function and significantly reduced microplastic translocation and liver damage, underscoring the importance of biomechanical cues in modeling systemic nanotoxicity.

Moreover, Yang et al. (2023) presented an integrated gut–liver-on-a-chip platform designed to model non-alcoholic fatty liver disease (NAFLD). The system dynamically simulated gut–liver interactions by fluidically linking Caco-2 and HepG2 cells under closed-loop perfusion, controlled by on-chip microvalves and a peristaltic micropump. This setup enabled precise, independent regulation of each chamber. The device featured a modified PDMS surface coated with DDM (n-dodecyl β-D-maltosid) and Matrigel to reduce compound absorption and enhance molecular stability. Physiological shear stress supported epithelial viability, while high-content imaging and transcriptomic analysis provided real-time, multi-layered insights into cellular responses. Kida et al. (2024) developed an all-PDMS gut-liver device integrating a paracellular permeability assay, enabling real-time visualization of epithelial transport influenced by hepatocyte excretions. In a toxicological application, Yu et al. (2024) simulated acetaminophen-induced liver injury on a gut-liver chip, observing mitochondrial stress, apoptosis markers, and ROS upregulation, confirming the model's validity for acute hepatotoxicity testing. In the same year, Wang et al. (2024c) introduced a Perfluoropolyether-based system to reduce drug sorption and improve first-pass metabolism measurement, allowing dissection of intestinal vs hepatic metabolism using genome-edited Caco-2 cells and primary hepatocytes. As toxicity often involves immune crosstalk, Jeon et al. (2020) developed a three-tissue MPS integrating gut (Caco-2), liver (HepG2), and immune (RAW264.7 macrophages) cells, showing that lipopolysaccharide-stimulated gut permeability triggers hepatic inflammation via nitric oxide and cytokine upregulation, positioning this model for systemic inflammatory toxicology. Similarly, Chen et al. (2017) reported that inflammatory mediators from gut-liver interactions suppressed hepatic expression (CYP7A1), supporting the system's application for evaluating inflammatory modulation of xenobiotic metabolism. These immune-enhanced systems were further evolved by Kim et al. (2025), who modeled NAFLD progression using a modular gut-liver-immune chip capable of recapitulating leaky gut-induced hepatic steatosis and cytokine-driven damage, confirming its utility for studying chronic dietary exposure effects. Beyond host tissues, Lucchetti et al. (2024b) developed a modular gut–microbiota–liver-on-a-chip system with bidirectional interconnection to study the microbiome-mediated reactivation of irinotecan, a chemotherapy drug associated with gastrointestinal toxicity. The setup linked two existing chips: (1) a gut-on-a-chip modeling the intestinal barrier and microbial activity, and (2) a liver-on-a-chip replicating hepatic architecture with key liver cell types, which responded to the reactivated metabolite. The study underscores the importance of incorporating gut microbiota into toxicological assessments, particularly for compounds undergoing enterohepatic recirculation or microbiome-dependent metabolism.

Extending beyond the liver, gut-kidney axis models are increasingly recognized for their importance in NP clearance, particularly due to the kidneys' vulnerability to nephrotoxicity resulting from NP accumulation in renal tubules. Even if less gut-kidney axis than liver axis can be found in the literature, still progress has been made in modeling the gut-kidney axis using organ-on-a-chip technologies. For instance, Lee et al. (2021) presents a Gut–Kidney Axis on a chip, engineered to model the inter-organ interactions between gut and kidney tissues under Shiga toxin-producing Escherichia coli (STEC) infection. Technically, the system integrates gut (Caco-2) and kidney (HKC-8) cell cultures within a modular, tiltable microfluidic chip, allowing for dynamic media perfusion. Complementing this, Li et al. (2017) showed that intestinal absorption modulators such as verapamil and colestyramine altered renal cell viability and lactate dehydrogenase (LDH) release, establishing a direct link between oral exposure and nephrotoxic outcomes. These studies highlight gut–kidney communication in nanotoxicology, but current models lack 3D nephron structures essential for capturing chronic renal toxicity. Further upstream, the gut–brain axis has opened avenues for neurotoxicity modeling. Alam et al. (2024) lay the conceptual foundation by mapping the complex biological pathways, neural, immune, endocrine, and microbial, through which the gut communicates with the brain. Their review highlights how emerging 3D in vitro models and microfluidic systems may overcome the limitations of traditional animal studies. As a recent concrete example, Jones et al. (2024) developed a dual-compartment gut–brain model consisting of two fluidically independent yet biologically interconnected module, one mimicking the colonic epithelium and the other comprising dopaminergic neuronal cultures derived from human cells. Moreover, Seo et al. (2024) contribute a microfluidic Gut-Brain-Axis tailored for real-time visualization of nanoscale vesicle transport using fluorescence-labeled exosomes. Its controlled flow dynamics and collagen-based ECM enable quantification of vesicle migration across barriers. Advancing biological relevance further, Kim et al. (2024b) use iPSC-derived neurons to examine the effects of microbial metabolites and extracellular vesicles on neural development and synaptic activity. Their chip-based model demonstrates how gut-derived signals influence neurotrophic pathways and neuronal differentiation within a physiologically controlled gut-brain environment. . These advancements in dual-organ gut-axis modeling reflect a broader evolution in MOC technologies. One of the pioneering works in the field was led by Maschmeyer et al. (2015), who introduced a foundational four-organ MPS, marking a significant milestone in MOC technology. Their platform interconnected intestine, liver, skin, and kidney models via integrated microfluidic circuits designed to sustain organ viability and enable dynamic inter-organ communication for up to 28 days. Key innovations included two distinct fluidic loops, one simulating systemic circulation and another mimicking renal excretion, alongside 3D liver spheroids and monolayers of kidney epithelial cells, enhancing physiological realism. On-chip peristaltic micropumps provided pulsatile flow to replicate in vivo-like shear stress, while a kidney membrane ensured directional metabolite transport through spatiotemporal fluid separation. Despite its strengths, the system's reliance on PDMS and customized inserts presented challenges for broader scalability and reproducibility. Trapecar et al. (2021) expanded the organ-on-chip paradigm by integrating immune and neurological elements within a mesofluidic platform. This system fluidically connected gut, liver, and cerebral MPS, incorporating circulating CD4+ regulatory T (Treg) and T helper 17 (TH17) immune cells to model the microbiome–gut–brain–immune axis. Constructed entirely from polycarbonate, the system integrated immune-compatible micropumps and offered scalable flow control. However, its complexity, limited systemic scope, and short co-culture duration constrained both its throughput and utility for systemic modeling. In parallel, Vernetti et al. (2017) proposed a modular strategy emphasizing "functional coupling" over direct fluidic integration. Their approach sequentially transferred metabolized media across five independently optimized human organ models-intestine, liver, kidney, BBB, and skeletal muscle, allowing precise evaluation of compound-specific ADME (absorption, distribution, metabolism, excretion) profiles and toxicity. Their studies with terfenadine, vitamin D3, and the microbiome-derived metabolite trimethylamine confirmed physiological relevance, notably showing that its hepatic metabolite could cross the BBB. The modular format enabled media-specific optimization and operational flexibility, yet lacked real-time feedback between organs and was challenged by media compatibility and flow dynamics. Extending systemic modeling, Koning et al. (2021) developed a multi-organ-on-chip system to investigate systemic immunotoxicity by connecting reconstructed human gingiva and skin tissues with Langerhans cells under dynamic flow, simulating how oral exposure to nickel sulfate can trigger immune responses in distant epithelial tissues. While no significant histological damage or cytokine release was observed, gene expression revealed increased immune activation markers, indicating Langerhans cell maturation and migration. This study was among the first to demonstrate cross-tissue immune signaling in a dynamic chip model, though limited to two epithelial tissues and a 72-hour duration. Recently, Fanizza et al. (2025) presented a landmark five-organ human-based platform incorporating the gut, immune system, liver, blood-brain barrier (BBB), and brain, each constructed from human primary or iPSC-derived cells and linked by a recirculating fluidic circuit that replicates physiological flow. The system models oral drug administration and systemic transport of the Alzheimer's drug donepezil, integrating computational simulations with functional biomarkers.

These studies demonstrated the feasibility of capturing multi-organ interactions in vitro. By incorporating dynamic features such as peristalsis (Wang et al., 2024c), immune responsiveness (Jeon et al., 2020; Kim et al., 2025), and microbiota-derived metabolite interactions (Lucchetti et al., 2024b; Kang et al., 2023), these models better reflect real-world exposure scenarios. Foundational work has paved the way for increasingly integrated systems, including three-organ configurations (Trapecar et al., 2021; Vernetti et al., 2017) and four to five-organ platforms (Maschmeyer et al., 2015; Koning et al., 2021, Fanizza et al., 2025). Despite these advances, several limitations remain. Many systems continue to rely on immortalized cell lines (e.g., Caco-2, HepG2), which lack the full metabolic and immunological capabilities of primary or iPSC-derived tissues.

From a technical standpoint, a major challenge in multi-organ integration is the need for organ-specific culture media. While tailored media are essential for maintaining distinct cellular functions, using a shared medium across modules can compromise cell viability and disrupt paracrine signaling. A practical solution involves compartmentalized chambers linked by semi-permeable membranes, which allow molecular exchange while keeping media isolated. For instance, Lucchetti et al. (2024) implemented this strategy by connecting four gut chambers and a liver module via nano/micro-porous membranes, enabling diffusion of signaling molecules without bulk media mixing. However, this setup still lacks precise control over the timing and concentration of exchanged signals, limiting its ability to reproduce dynamic and directional gradients seen in vivo.

Another critical challenge is maintaining physiologically relevant oxygen and nutrient gradients across compartments, given the distinct metabolic needs of each organ. In shared perfusion systems, downstream tissues often receive non-physiological concentrations of oxygen or nutrients. Jiang et al. (2024b) addressed this by designing a closed-loop platform with integrated oxygen control and inline sensors, enabling real-time adjustments of oxygen levels (4–20%) and nutrient delivery tailored to each module's requirements.

Moreover, fluidic directionality remains a key hurdle. Bidirectional flow is essential to replicate inter-organ communication via hormones, cytokines, and metabolites feedback that unidirectional systems fail to capture. Without reciprocal signaling, upstream organs lack critical physiological input. Maschmeyer et al. (2015) addressed this by incorporating dual peristaltic pumps to drive pulsatile, bidirectional flow through a four-organ system. Similarly, Lucchetti et al. (2024) used a dual-loop configuration with porous membranes to enable two-way diffusion between gut and liver compartments. Still, both approaches face limitations in controlling the timing of feedback signals, which is essential for faithfully mimicking the temporal dynamics of systemic physiology.

Long-term culture stability is also often insufficient for modeling chronic dietary exposures. Furthermore, only a limited number of models replicate key aspects of the gut lumen, such as mucus layers, microbial populations, and food matrix interactions, that critically influence NP behavior. Finally, challenges in scalability and throughput hinder broader regulatory adoption, which requires reproducible data across a wide range of compounds and exposure scenarios.

### 2.5. Recent Advances in Gut-on-Chip Technologies for Nanotoxicology Applications

As previously discussed, state-of-the-art MPS, including GOC and MOC platforms, offer powerful tools for nanotoxicity assessment. Nevertheless, their application in this context remains limited. This section presents representative case studies illustrating the application of GOC in nanotoxicity assessment.

The transition from conventional 2D monolayers to 3D gut models has been a crucial step in nanotoxicology. Chia et al. (2015) employed a biomimetic 3D gastrointestinal spheroid platform to evaluate the inflammatory effects of ZnO-NPs. Their study found that ZnO-NPs triggered an inflammatory response in 3D spheroids, suggesting that these models may be better suited for toxicological assessments than traditional 2D cultures. Expanding on this, Wu et al. (2017) compared the toxicity of different-sized ZnO-NPs in both 2D monolayers and 3D gut cultures. Their findings indicated that 3D cultures were more resistant to NP-induced toxicity than monolayers, reinforcing the idea that traditional 2D models may overestimate cytotoxicity. Sokolova et al. (2022) investigated ultrasmall Au-NPs using a 3D gut epithelial model composed of Caco-2 and human colon (HT29-MTX) cells. The study revealed that while these NPs were efficiently taken up by cells, their toxicity was minimal. This reinforces the importance of using physiologically relevant models to differentiate between toxic and non-toxic NPs. Similarly, Gautam et al. (2024) compared NP toxicity in cell monolayers versus 3D gut epithelial cultures, demonstrating that 3D models provide a more physiologically relevant assessment of NP interactions with the gut lining. Their study demonstrated that 3D gut epithelial cultures exhibited different cellular responses compared to monolayers, emphasizing the need for advanced models in food nanotoxicology.

One of the key strengths of GOC models lies in their ability to simulate and measure NP transport across the intestinal barrier. Pocock et al. (2019) developed an intestine-on-a-chip model that recreated a mucus-secreting epithelial barrier using Caco-2 cells under fluid shear. The design enabled real-time imaging of NP interactions with mucus and epithelium. Using PEGylated and non-PEGylated silica NPs of varying sizes, the study showed enhanced uptake and mucus penetration with PEG (polyethylene glycol) coatings. Inversion experiments confirmed sedimentation effects on larger particles. The platform provided a physiologically relevant and accessible system for studying NP transport and mucosal delivery. Additionally, Jia et al. (2021) developed a Mucus-on-Chip platform to analyze NP penetration into mucus, an important barrier affecting drug and toxin absorption. This study demonstrated that nanoparticulate drug carriers interact dynamically with mucus, a factor often overlooked in static cell cultures. However, the model lacked peristalsis and a complex cellular environment, restricting its ability to fully replicate human gut physiology. To better replicate gut dynamics, Santoni et al. (2025) introduced a co-culture GOC model with orbital mechanical stimulation, simulating intestinal motility to study the impact of food-relevant NPs. This study demonstrated that mechanical forces influence epithelial responses to NPs, highlighting the need to incorporate biomechanical cues in toxicity assessments. Delon et al. (2023) used a microphysiological intestine-on-a-chip model to study the transcytosis of polystyrene NPs across Caco-2 cell monolayers. The model incorporated fluid flow to mimic intestinal microcirculation, offering insights into real-time particle capture and quantification. Wright et al. (2023) further contributed to this area by designing a membrane-free microfluidic system to distinguish between mucoadhesive and mucopermeating NPs. This approach improved the efficiency of NP screening for drug delivery but did not assess their potential cytotoxicity or inflammatory effects, which are critical factors in nanotoxicology.

Han et al. (2025) further explored the role of peristalsis in GOC, showing that enhanced peristalsis reduced inflammation induced by nanoplastics. This suggests that biomechanical forces play a protective role in intestinal homeostasis, a factor that static models fail to capture. This study represents a step forward in using GOC platforms for nanotoxicity assessment, though it did not include chronic exposure assessments, which remain a limitation in the field. Although GOC models have been used in nanotoxicity studies, their potential remains underexploited, as most lack microbiota and immune cell integration, key components that modulate nanoparticle interactions in vivo. The table 4 compares the main characteristics of the latest advanced in GOC with those of GOC developed for nanotoxicological applications. While state-of-the-art models increasingly replicate the complexity of the human intestinal microenvironment, including immune, microbial, and mechanical components, nanotoxicology-focused systems often remain simplified, prioritizing analytical throughput and barrier function assessment. The comparison highlights a notable gap between the technological maturity of GOC systems and their underutilization in the field of nanotoxicology.

Table 4: Comparative Summary: State-of-the-Art vs Nanotoxicology Gut Models

| Category | Studies | Focus | Stimulation | Strengths | Limitations |
|---|---|---|---|---|---|
| State-of-the-Art Models | Zhang et al., 2024; Ballerini et al., 2025; Kim et al., 2024a; De Gregorio et al., 2022; | Host–microbiota–immune interactions, epithelial physiology, disease modeling | Peristalsis, flow, $O_2$ gradients | High fidelity, immune and microbial co-culture, dynamic microenvironments, predictive of host response | Complex, resource-intensive, lower throughput for toxicity |
| Nanotoxicology-Oriented Models | Santoni et al., 2025; Sokolova et al., 2022; Wright | Barrier integrity, NP uptake, mucus interaction, | Flow, peristalsis-like strain, orbital shear | High-throughput compatible, quantitative mucopenetration | Reduced immunological and microbial complexity, |

| | et al., 2023; Wu et al., 2017; Gautam et al., 2024; Jia et al., 2021; Delon et al., 2023; Han et al., 2025 | oxidative stress, dynamic inflammation detection | | , mechanistic resolution of NP impact | limited longitudinal host interaction modeling |

The comparative analysis of GOC systems clearly illustrates two diverging trajectories: one driven by the pursuit of physiological realism, and the other by the need for toxicological precision. On one hand, state-of-the-art platforms as exemplified by studies such as Zhang et al. (2024), Ballerini et al. (2025), Kim et al. (2024a) and De Gregorio et al. (2022) have pushed the boundaries of microphysiological modeling. These systems integrate key intestinal features, including anaerobic microbiota, adaptive immune components, mucus-producing cells, and mechanical cues such as peristalsis and oxygen gradients, enabling highly relevant investigations of host–microbiota–immune crosstalk, mucosal inflammation, and disease-specific mechanisms. In contrast, GOC designed for nanotoxicology, including recent studies by Han et al. (2025), Jia et al. (2021), Delon et al. (2023), and Gautam et al. (2024), have remained comparatively modest in complexity. While these models often incorporate important features such as mucus barriers, peristalsis-like strain, and real-time biosensors, they typically lack integration of immune components and microbiota, which are critical modulators of NP fate and toxicity in vivo. Their design emphasizes mechanistic clarity and throughput, which suits regulatory screening but underutilizes the broader advances made in GOC engineering.

This disparity highlights a key observation: the field of GOC for nanotoxicology remains underdeveloped relative to the overall technological progress of the domain. While tremendous strides have been made in replicating the human intestinal microenvironment for biomedical applications, these advances have not yet been fully leveraged to investigate NP-induced toxicity, particularly in the context of host–microbe–immune interactions. Bridging this gap presents a critical opportunity to transform nanotoxicology, enabling more predictive, mechanistic, and human-relevant assessments of nano–bio interactions, particularly in the emerging field of nano-enabled foods.

### 3. Challenges and Future Perspectives in Nanofood Safety: Tipping the Scales with MPS

Recent advances in MPS technologies have led to increasingly physiologically relevant gut models and multi-organ platforms that simulate interactions between the gut and other tissues. These innovations offer a promising alternative for assessing the toxicity of nanomaterials in food. To date, most progress in MPS has been driven by the pharmaceutical sector, motivated by the need for more predictive drug testing models and adherence to the 3Rs principle (Reduction, Refinement, and Replacement of animal testing). As this field continues to evolve, even more sophisticated systems are expected to emerge. However, translating these models to food safety poses unique challenges. Unlike pharmaceuticals where therapeutic benefits can tip the scales in favor of accepting certain risks, nanomaterials in food, used primarily to enhance texture, stability, or nutrient delivery, are often seen as non-essential, offering modest benefits that carry less weight. As a result, even minor signs of toxicity, bioaccumulation, or immune activation can tip the balance toward public concern, regulatory scrutiny, or outright rejection. The threshold for demonstrating safety in food applications is therefore significantly higher. Applying MPS systems in this context demands careful consideration of sector-specific expectations and risk perceptions. While MPS platforms will benefit from ongoing pharmaceutical advancements, tailored adaptations are essential to meet the distinct demands of food-related nanotoxicology. This section outlines future strategies to advance MPS in food nanotoxicology, emphasizing the need for realistic exposure models and context-specific refinements both essential to rebalance the scales, reduce perceived risks, and build public and regulatory confidence in nano-enabled food products.

### 3.1. Realistic Exposure Pathways and Chronic Dose–Effect Relationships

Dietary nanomaterials are typically encountered at low concentrations over extended periods, highlighting the need for long-term culture systems that simulate chronic exposure and reflect real-world ingestion patterns. Such systems are essential for investigating cumulative effects like nanoparticle (NP) bioaccumulation, biotransformation, and low-grade inflammation, and are critical for detecting delayed physiological outcomes, including immune modulation, barrier dysfunction, and microbial dysbiosis.

To ensure physiological relevance, future MPS platforms should be designed to:
- Maintain stable, long-term co-cultures of intestinal epithelium, immune cells, and microbiota, still limited by technical complexity.
- Use NP concentrations reflecting actual dietary exposure.
- Simulate oral ingestion, including enzymatic digestion and gastric preconditioning, prior to intestinal contact.
- Account for biotransformation within the gastrointestinal environment, where NP behavior may shift through interactions with enzymes, mucus, or microbial metabolites.
- Integrate multi-organ systems to track NP ADME. These models enable study of systemic effects such as hepatotoxicity, immune activation, and metabolite formation, offering a more complete risk profile.

By aligning experimental designs with human physiology and dietary habits, MPS can deliver more actionable, regulatory-relevant insights into nanomaterial safety.

### 3.2. Standardization and Validation of MPS Models

While the innovation in OoC technology has been remarkable, it has also led to a highly fragmented landscape. Variations in chip materials, flow dynamics, cell sources, and analytical endpoints make it difficult to compare results across laboratories, undermining reproducibility and slowing regulatory acceptance.

To fully harness the potential of MPS in food nanotoxicology, standardization efforts must prioritize:
- Harmonized protocols for exposure conditions, biological markers, and analytical techniques.
- Enhanced scalability and reproducibility to support routine use in both academic and industrial settings.

- Integration of real-time monitoring systems such as sensors for ROS, inflammatory cytokines, and microbiome shifts, to enable dynamic and mechanistic assessments.
- Development of user-friendly and cost-efficient platforms, expanding accessibility beyond specialized research labs.

While MPS adoption is advancing in the pharmaceutical industry, driven by the need for better preclinical tools and reduced animal testing, the agri-food sector remains behind, constrained by fewer validated models.

### 3.3. Artificial Intelligence: A Critical Complement to MPS

As MPS become increasingly complex and data-rich, the integration of artificial intelligence (AI) tools such as in silico modeling and machine learning algorithms, is becoming essential. AI offers powerful capabilities for analyzing high-dimensional, time-resolved datasets produced by organ-on-chip experiments, identifying subtle toxicity signals that may be missed through conventional analysis. AI can further strengthen MPS-based nanotoxicology by:

- Detecting early biomarkers of adverse responses based on temporal cytokine shifts, barrier function degradation, or microbial imbalances.
- Simulating extended exposure scenarios, extrapolating from limited experimental timeframes to longer-term outcomes.
- Assisting in experimental design optimization, including dose selection, sampling intervals, and endpoint prioritization.
- Supporting the integration of gut and other organ modules by modeling optimal flow dynamics, predicting inter-organ signaling patterns, and guiding microfluidic design choices that preserve physiological connectivity.

Beyond these applications, AI can also contribute significantly to standardization efforts by identifying reproducible patterns across diverse MPS setups, informing protocol harmonization, and facilitating cross-study comparisons. This synergy between AI and MPS not only enhances predictive accuracy but also supports the development of robust, scalable models, laying the groundwork for personalized risk profiling and precision food safety assessment..

### 3.4. Bridging the Gap: Collaborative Innovation for Food-Specific MPS

Despite their technical promise, MPS remain underutilized in food safety assessment. Most development and funding focus on biomedical applications, while the food sector lacks the interdisciplinary collaboration and infrastructure needed to tailor these platforms for dietary nanotoxicology. To overcome this, there is a need for collaborative and cross-disciplinary initiatives. Academic researchers, food scientists, toxicologists, regulatory bodies, and industry stakeholders must co-develop MPS models that are fit for purpose in food safety assessment. Strategic goals should include:

- Establishing shared standards and reference datasets.
- Creating open-access databases for benchmarking and model validation.
- Engaging with regulators early in the development process to align models with evolving safety assessment criteria.

By uniting technological innovation with regulatory alignment and open collaboration, the food sector can fully harness the potential of MPS to advance nanomaterial safety. This coordinated approach is essential for ensuring science-driven decision-making in future food systems.

**Conclusion**

Gut-focused MPS platforms offer a transformative tool for assessing the safety of ingested nanomaterials in food. When tailored to food nanotoxicology and integrated with emerging technologies from Industry 4.0, 5.0 and the emerging vision of Industry 6.0, including AI, precision microfabrication, real-time biosensing, and automation, MPS can help unlock the potential of nano-enabled food products. This integration has the potential to improve the predictive power and efficiency of safety assessments while supporting the development of safer, more sustainable food systems aligned with SDGs 2, 3, and 12.
While gut-MPS platforms are well-suited for evaluating ingested nano-additives, they are not sufficient for comprehensive risk assessments involving food packaging, where exposure may also occur through dermal contact or inhalation. MPS platforms mimicking skin and lung environments can address these non-oral pathways, enabling more holistic safety evaluations. Yet scientific innovation alone is not enough. In the food sector, where consumer trust is fragile and perceived benefits may be incremental, the balance between innovation and safety must be carefully maintained. MPS technologies, if designed collaboratively and validated rigorously, can serve as precision tools on this balance scale, adding weight to the safety side through realism, transparency, and standardization. With continued investment in interdisciplinary collaboration, infrastructure, and regulatory alignment, MPS platforms can ensure that food nanotechnologies are not only cutting-edge but also safe, trusted, and socially responsible. Ultimately, by fostering public confidence and enabling broader access to safe technologies, these systems can contribute to healthier communities, more sustainable food production, and a fairer distribution of innovation benefits across society.

**Acknowledgments**

This review article was made possible through the support of several key initiatives. Expertise in the development and application of Organ-on-a-Chip technologies was supported through the CISTEM project that received funding from the European Union's Horizon 2020 research and innovation programme under the Marie Skłodowska-Curie Grant Agreement No. 778354. In parallel, insights into the use of nanomaterials in food applications particularly regarding their functionality, behavior, and potential safety implications, were supported through the ANTARES project that received funding from the European Union's Horizon 2020 research and innovation programme under Grant Agreement SGA-CSA No. 739570 under FPA. Additional support from the Ministry of Science, Technological Development and Innovation of the Republic of Serbia (Grant No. 2024: 451–03–66/2024–03/200358) further contributed to the advancement of this work.